\documentclass{PoS}

\usepackage{cite}
\usepackage{axodraw}
\usepackage{float}
\usepackage{shuffle}

\usepackage[utf8]{inputenc}

\newcommand{\bq}{\begin{eqnarray*}}
\newcommand{\eq}{\end{eqnarray*}}
\newcommand{\eps}{\varepsilon}

\title{Relations and representations of QCD amplitudes}

\ShortTitle{Relations and representations of QCD amplitudes}

\author{Leonardo de la Cruz\\
        PRISMA Cluster of Excellence, Johannes Gutenberg-Universit\"at Mainz\\
        E-mail: \email{ledelacr@uni-mainz.de}}

\author{Alexander Kniss\\
        PRISMA Cluster of Excellence, Johannes Gutenberg-Universit\"at Mainz\\
        E-mail: \email{akniss@students.uni-mainz.de}}

\author{\speaker{Stefan Weinzierl}\\
        PRISMA Cluster of Excellence, Johannes Gutenberg-Universit\"at Mainz\\
        E-mail: \email{weinzierl@uni-mainz.de}}

\abstract{
In this talk we review relations and representations of primitive QCD tree amplitudes.
Topics covered include the BCJ relations, the CHY representation, and the KLT relations.
We will put a special emphasis on how these relations and representations
generalise from pure Yang-Mills theory to QCD.
The generalisation of the KLT relations from pure Yang-Mills to QCD includes the case of massive quarks.
On the gravity side we then obtain hypothetical particles interacting with gravitational strength,
which can be massive and non-relativistic.
}

\FullConference{Loops and Legs in Quantum Field Theory\\
		24-29 April 2016\\
		Leipzig, Germany}

\begin{document}


\section{Introduction}

In this talk we discuss relations and representations of QCD amplitudes.
We will pay special attention to the question, which properties of amplitudes in pure Yang-Mills theory carry
over to QCD amplitudes.
Pure Yang-Mills theory is a theory of massless gauge bosons and nothing else.
The gauge bosons are in the adjoint representation of a Lie group $G$.
For a non-abelian gauge group this is an interesting theory due to the self-interactions of the massless gauge gluons.
We will call the gauge bosons gluons, but it should be noted that 
for the topics covered in this talk nothing will depend on the specific choice $G=\mathrm{SU}(3)$ for the gauge group.

QCD is a larger theory, where in addition to massless gauge bosons we have $N_f$ different flavours
of fermions in the fundamental representation of the gauge group.
We will call the fermions quarks.
Flavour is conserved in all interactions.
In contrast to the gauge bosons, which must be massless, the fermions may be massive.

In this talk we consider scattering amplitudes with $n$ external particles.
We assume that $n_g$ of these are gluons, $n_q$ are quarks and $n_q$ are anti-quarks.
We therefore have
\bq 
 n & = & n_g + 2 n_q.
\eq
We will focus on Born amplitudes.

QCD amplitudes may be decomposed into sums of
group-theoretical factors multiplied by kinematic functions called primitive amplitudes. 
For the $n$-gluon tree amplitude this decomposition reads
\bq
{\cal A}_{n}^{\mathrm{YM}}(1,2,...,n) & = & g^{n-2} \sum\limits_{\sigma \in S_{n}/Z_{n}} 
 2 \; \mbox{Tr} \left( T^{a_{\sigma(1)}} ... T^{a_{\sigma(n)}} \right)
 \;\;
 A_{n}^{\mathrm{YM}}\left( {\sigma(1)}, ..., {\sigma(n)} \right),
\eq
where the sum is over all non-cyclic permutations of the external legs.
The primitive amplitudes have several nice properties:
(i) By construction, all group-theoretical factors have been stripped off.
(ii) The primitive amplitudes are gauge invariant.
(iii) Each primitive amplitude has a fixed cyclic order of the external legs.
The colour decomposition into primitive amplitudes can be constructed algorithmically
for all tree and one-loop 
QCD amplitudes \cite{Ellis:2008qc,Ellis:2011cr,Ita:2011ar,Badger:2012pg,Reuschle:2013qna,Schuster:2013aya}.
The primitive amplitudes are calculated from cyclic-ordered Feynman rules:
The rules for the propagators are
\bq
\begin{picture}(85,20)(0,5)
 \ArrowLine(70,10)(20,10)
\end{picture} 
 \;\; = \;\;
 i\frac{p\!\!\!/+m}{p^2-m^2},
 & &
\begin{picture}(85,20)(0,5)
 \Gluon(20,10)(70,10){-5}{5}
\end{picture} 
 \;\; = \;\;
 \frac{-ig^{\mu\nu}}{p^2}.
\eq
The cyclic-ordered Feynman rules for the three-gluon and the four-gluon vertices are
\bq
\begin{picture}(100,35)(0,50)
\Vertex(50,50){2}
\Gluon(50,50)(50,80){3}{4}
\Gluon(50,50)(76,35){3}{4}
\Gluon(50,50)(24,35){3}{4}
\LongArrow(56,70)(56,80)
\LongArrow(67,47)(76,42)
\LongArrow(33,47)(24,42)
\Text(60,80)[lt]{$p_1^{\mu_1}$}
\Text(78,35)[lc]{$p_2^{\mu_2}$}
\Text(22,35)[rc]{$p_3^{\mu_3}$}
\end{picture}
 & = &
 i \left[ g^{\mu_1\mu_2} \left( p_1^{\mu_3} - p_2^{\mu_3} \right)
         +g^{\mu_2\mu_3} \left( p_2^{\mu_1} - p_3^{\mu_1} \right)
         +g^{\mu_3\mu_1} \left( p_3^{\mu_2} - p_1^{\mu_2} \right)
   \right],
 \nonumber \\
 \nonumber \\
 \nonumber \\
\begin{picture}(100,35)(0,50)
\Vertex(50,50){2}
\Gluon(50,50)(71,71){3}{4}
\Gluon(50,50)(71,29){3}{4}
\Gluon(50,50)(29,29){3}{4}
\Gluon(50,50)(29,71){3}{4}
\Text(72,72)[lb]{\small $\mu_1$}
\Text(72,28)[lt]{\small $\mu_2$}
\Text(28,28)[rt]{\small $\mu_3$}
\Text(28,72)[rb]{\small $\mu_4$}
\end{picture}
 & = &
  i \left[
        2 g^{\mu_1\mu_3} g^{\mu_2\mu_4} - g^{\mu_1\mu_2} g^{\mu_3\mu_4} 
                                        - g^{\mu_1\mu_4} g^{\mu_2\mu_3}
 \right].
 \nonumber \\
 \nonumber \\
\eq
The Feynman rule for the quark-gluon vertex is given by
\bq
\label{colour_ordered_quark_gluon_antiquark_vertex}
\begin{picture}(100,35)(0,50)
\Vertex(50,50){2}
\Gluon(50,50)(80,50){3}{4}
\ArrowLine(50,50)(29,71)
\ArrowLine(29,29)(50,50)
\Text(82,50)[lc]{$\mu$}
\end{picture}
 \;\; = \;\;
 i \gamma^{\mu},
 & \;\;\;\;\;\;\;\;\; &
\begin{picture}(100,35)(0,50)
\Vertex(50,50){2}
\Gluon(50,50)(20,50){3}{4}
\ArrowLine(50,50)(71,71)
\ArrowLine(71,29)(50,50)
\Text(18,50)[rc]{$\mu$}
\end{picture}
 \;\; = \;\;
 -i \gamma^{\mu}.
 \nonumber \\
 \nonumber \\
\eq
In practice, primitive QCD amplitudes (and full QCD amplitudes)
are very efficiently calculated with the help of Berend-Giele
recurrence relations \cite{Berends:1987me,Dinsdale:2006sq,Duhr:2006iq}.
From a practical phenomenology-oriented point of view this is all what we need
and we might be inclined to consider QCD primitive tree amplitudes a settled case.
However, there is more structure hidden in QCD primitive tree amplitudes, which is not evident in a Feynman diagram based
calculation or an approach based on recurrence relations.
This talk is about these additional structures.
Most of these additional structures have been first discussed in the context of pure Yang-Mills theory.
We will put a special emphasis on how these structures generalise from pure Yang-Mills theory to QCD.


\section{BCJ relations}

An obvious question related to the colour decomposition is: 
How many independent primitive amplitudes are there for $n$ external particles?
Let us first look at the pure Yang-Mills case.
Primitive amplitudes are labelled by a permutation, specifying the order of the external particles.
For $n$ external particles there are $n!$ permutations and therefore $n!$ different orders.
However, there are relations among primitive amplitudes with different external order.
The first set of relations is given by cyclic invariance:
\bq
 A_n^{\mathrm{YM}}(1,2,...,n) & = & A_n^{\mathrm{YM}}(2,...,n,1)
\eq
Cyclic invariance is the statement that only the cyclic external order matters, not the point, where we start to read
of the order.
Cyclic invariance reduces the number of independent primitive amplitudes to $(n-1)!$.

The first non-trivial relations are the Kleiss-Kuijf relations \cite{Kleiss:1988ne}, 
which follow from the anti-symmetry of the colour-stripped three-valent vertices.
To state the Kleiss-Kuijf relations we set
$\vec{\alpha} = (\alpha_1, ..., \alpha_j)$, $\vec{\beta} = (\beta_1, ..., \beta_{n-2-j})$
and $\vec{\beta}^T = (\beta_{n-2-j}, ..., \beta_1)$.
The Kleiss-Kuijf relations read
\bq
 A_n^{\mathrm{YM}}( 1, \vec{\beta}, 2, \vec{\alpha} )
 & = & 
 \left( -1 \right)^{n-2-j}
 \sum\limits_{\sigma \in \vec{\alpha} \; \shuffle \; \vec{\beta}^T}
 A_n^{\mathrm{YM}}( 1, 2, \sigma_1, ..., \sigma_{n-2} ).
\eq
Here, $\vec{\alpha} \; \shuffle \; \vec{\beta}^T$ denotes the set of all shuffles of $\vec{\alpha}$ with $\vec{\beta}^T$, i.e.
the set of all permutations of the elements of $\vec{\alpha}$ and $\vec{\beta}^T$, which preserve the relative order of the
elements of $\vec{\alpha}$ and of the elements of $\vec{\beta}^T$.
The Kleiss-Kuijf relations reduce the number of independent primitive amplitudes to $(n-2)!$.

Finally, we have the fundamental Bern-Carrasco-Johansson relations (BCJ relations) \cite{Bern:2008qj,Feng:2010my}.
\bq
 \sum\limits_{i=2}^{n-1} 
  \left( \sum\limits_{j=i+1}^n 2 p_2 p_j \right)
  A_n^{\mathrm{YM}}(1,3,...,i,2,i+1,...,n-1,n)
 & = & 0.
\eq
These reduce the number of independent primitive amplitudes to $(n-3)!$.
The full set of relations among primitive tree amplitudes in pure Yang-Mills theory is given by cyclic invariance,
Kleiss-Kuijf relations, and the fundamental BCJ relations.
Therefore a basis of independent primitive amplitudes consists of $(n-3)!$ elements.

The BCJ relations are closely linked to Jacobi-like identities for kinematical numerators.
This is known under the name ``colour-kinematics duality'' \cite{Bern:2010ue} 
and states that Yang-Mills amplitudes can be brought into a form
\bq
 {\mathcal A}_n^{\mathrm{YM}}
 & = &
 i g^{n-2}
 \sum\limits_{\mathrm{trivalent} \; \mathrm{graphs} \; G}
 \frac{C\left(G\right)N\left(G\right)}{D\left(G\right)},
 \;\;\;\;\;\;\;\;\;\;\;\;
 D\left(G\right) = \prod\limits_{\mathrm{edges} \; e} s_e,
\eq
where the kinematical numerators $N(G)$ satisfy Jacobi-like relations, whenever the corresponding
colour factors $C(G)$ do:
\bq
\begin{picture}(100,50)(0,25)
\Vertex(50,30){2}
\Vertex(35,45){2}
\Line(50,5)(50,30)
\Line(50,30)(80,60)
\Line(50,30)(20,60)
\Line(35,45)(50,60)
\Text(20,65)[b]{\footnotesize $1$}
\Text(50,65)[b]{\footnotesize $2$}
\Text(80,65)[b]{\footnotesize $3$}
\Text(50,0)[t]{\footnotesize $4$}
\end{picture}
 +
\begin{picture}(100,50)(0,25)
\Vertex(50,30){2}
\Vertex(35,45){2}
\Line(50,5)(50,30)
\Line(50,30)(80,60)
\Line(50,30)(20,60)
\Line(35,45)(50,60)
\Text(20,65)[b]{\footnotesize $2$}
\Text(50,65)[b]{\footnotesize $3$}
\Text(80,65)[b]{\footnotesize $1$}
\Text(50,0)[t]{\footnotesize $4$}
\end{picture}
 +
\begin{picture}(100,50)(0,25)
\Vertex(50,30){2}
\Vertex(35,45){2}
\Line(50,5)(50,30)
\Line(50,30)(80,60)
\Line(50,30)(20,60)
\Line(35,45)(50,60)
\Text(20,65)[b]{\footnotesize $3$}
\Text(50,65)[b]{\footnotesize $1$}
\Text(80,65)[b]{\footnotesize $2$}
\Text(50,0)[t]{\footnotesize $4$}
\end{picture}
 & = & 0
 \nonumber \\
\eq
\bq
 C\left(G_1\right) + C\left(G_2\right) + C\left(G_3\right) = 0
 \hspace*{5mm}
 & \Rightarrow &
 \hspace*{5mm}
 N\left(G_1\right) + N\left(G_2\right) + N\left(G_3\right) = 0
 \hspace*{6mm}
\eq
Let us now consider primitive tree amplitudes in QCD.
Again, we may ask how many independent primitive amplitudes are there and what are the relations?
As before we have the trivial relations of cyclic invariance and the Kleiss-Kuijf relations.
In addition, we now have no-crossed-fermion-lines relations
\bq
\label{no_crossed_fermions}
 A_n\left( ..., i_q, ..., j_{q'}, ..., k_{\bar{q}}, ..., l_{\bar{q}'}, ... \right)
 \;\; = \;\;
 A_n\left( ..., i_q, ..., j_{\bar{q}'}, ..., k_{\bar{q}}, ..., l_{q'}, ... \right)
 \;\; = \;\;
 0,
\eq
which follow from the fact that flavour is conserved in QCD.
These relations may be used to orient the fermion lines \cite{Melia:2013bta,Melia:2013epa}.
Due to these no-crossed-fermion-lines relations we cannot have the full set of BCJ relations.
The set of fundamental BCJ relations for primitive QCD amplitudes is given by
\bq
 \sum\limits_{i=2}^{n-1} 
  \left( \sum\limits_{j=i+1}^n 2 p_2 p_j \right)
  A_n^{\mathrm{QCD}}(1,3,...,i,2_{g},i+1,...,n-1,n)
 & = & 0,
\eq
where particle $2$ is required to be a gluon.
These relations have been first conjectured in \cite{Johansson:2015oia}
and have been proven in \cite{delaCruz:2015dpa}.
This gives us the size of a basis of independent amplitudes:
\bq
 N_{\mathrm{basis}}
 & = &
 \left\{
 \begin{array}{ll}
   \left(n-3\right)!, & n_q \in \{0,1\}, \\
   \left(n-3\right)! \frac{2\left(n_q-1\right)}{n_q!}, & n_q \ge 2 \\
 \end{array}
 \right.
\eq
To give an example, we have four independent primitive tree amplitudes for three quark pairs:
\bq
 A_6(q_1,q_2,q_3,\bar{q}_3,\bar{q}_2,\bar{q}_1),
 \;\; 
 A_6(q_1,q_3,\bar{q}_3,q_2,\bar{q}_2,\bar{q}_1),
 \;\; 
 A_6(q_1,q_3,q_2,\bar{q}_2,\bar{q}_3,\bar{q}_1),
 \;\; 
 A_6(q_1,q_2,\bar{q}_2,q_3,\bar{q}_3,\bar{q}_1)
\eq
In contrast, for six gluons there are six independent primitive tree amplitudes:
\bq
 A_6(g_1,g_2,g_3,g_4,g_5,g_6),
 \;\; 
 A_6(g_1,g_3,g_4,g_2,g_5,g_6), 
 \;\; 
 A_6(g_1,g_4,g_2,g_3,g_5,g_6), 
 \nonumber \\
 A_6(g_1,g_4,g_3,g_2,g_5,g_6), 
 \;\; 
 A_6(g_1,g_3,g_2,g_4,g_5,g_6), 
 \;\; 
 A_6(g_1,g_2,g_4,g_3,g_5,g_6)
\eq


\section{CHY representation}

Primitive tree amplitudes depend on the order of the external particles but also on the spin/ helicity configuration
of the external particles.
Within a Feynman diagrams based calculation or an approach based on recurrence relations, the information on the external
order is interwoven with the information on the spin/helicity configuration.
Can we separate the dependence on the external order from the dependence on the spin/helicity configuration?
Let us again first look at the pure Yang-Mills case.
The Cachazo-He-Yuan representation (CHY representation) \cite{Cachazo:2013gna,Cachazo:2013hca,Cachazo:2013iea} achieves this.
In order to present the CHY representation we slightly change the notation.
We denote the external order by a word $w=(\sigma_1,...,\sigma_n)$, 
the $n$-tuple of external momenta by $p=(p_1,...,p_n)$ and
the $n$-tuple of external polarisations by $\eps=(\eps_1,...,\eps_n)$.
The $n$-gluon tree amplitude 
has a representation in the form of a global residue \cite{Sogaard:2015dba,Bosma:2016ttj}:
\bq
 A_n^{\mathrm{YM}}\left(w, p, \eps \right)
 & = &
 i
 \sum\limits_{\mathrm{solutions} \; j} J\left(z^{(j)},p\right) \; C\left(w, z^{(j)}\right) \; E\left(z^{(j)},p,\eps\right)
\eq
The sum is over the inequivalent solutions $z=(z_1,z_2,...,z_n)$ of the scattering equations
\bq
 f_i\left(z,p\right) & = & 
 \sum\limits_{j=1, j \neq i}^n \frac{ 2 p_i \cdot p_j}{z_i - z_j}
 \;\; = \;\; 0.
\eq
The function $C\left(w, z^{(j)}\right)$ encodes the information on the external ordering,
the function $E\left(z^{(j)},p,\eps\right)$ encodes the information on the external polarisations,
and 
$J(z^{(j)},p)$ is a Jacobian factor. The explicit expressions for these functions can be found in \cite{Cachazo:2013hca}.

We may then ask if there is a similar representation for primitive QCD tree amplitudes, which separates
the information on the external order from the one on the polarisations?
This is indeed the case.
We first generalise the scattering equations to allow for masses \cite{Naculich:2014naa,delaCruz:2015raa}
\bq
 \hat{f}_i\left(z,p\right) & = & 
 \sum\limits_{j=1, j \neq i}^n \frac{ 2 p_i \cdot p_j + 2 \Delta_{ij}}{z_i - z_j}
 \;\; = \;\; 0.
\eq
with 
$\Delta_{q_a \bar{q}_a} = \Delta_{\bar{q}_a q_a} = m_{q_a}^2$ and $\Delta_{ij}=0$ in all other cases.
In order to keep the notation simple, we continue to denote the set of external polarisations by $\eps$,
consisting of polarisation vectors $\eps_j$ for external gluons,
spinors $\bar{u}_j$ for out-going fermions and spinors $v_j$ for out-going anti-fermions.
Primitive tree amplitudes in QCD have a CHY representation similar to the one in pure Yang-Mills theory \cite{Weinzierl:2014ava,delaCruz:2015raa}:
\bq
 A_n^{\mathrm{QCD}}\left(w, p, \eps \right)
 & = &
 i
 \sum\limits_{\mathrm{solutions} \; j} J\left(z^{(j)},p\right) \; \hat{C}\left(w, z^{(j)}\right) \; \hat{E}\left(z^{(j)},p,\eps\right)
\eq
The function $\hat{C}\left(w, z^{(j)}\right)$ can be constructed from the relations between
primitive QCD tree amplitudes, while the 
construction of $\hat{E}\left(z^{(j)},p,\eps\right)$ is based on pseudo-inverse matrices \cite{delaCruz:2015raa}.
We note that such a representation is not unique. One may always multiply $\hat{C}$ and divide $\hat{E}$ by
a function of cross-ratios of the variables $z$.


\section{KLT relations}

The Kawai-Lewellen-Tye (KLT) relations \cite{Kawai:1985xq} provide a relation between primitive tree amplitudes
in pure Yang-Mills theory and graviton amplitudes in perturbative quantum gravity.
Within perturbative quantum gravity one considers (small) fluctuations around the flat Minkowski metric 
\bq
 g_{\mu\nu} & = &
 \eta_{\mu\nu} + \kappa h_{\mu\nu},
\eq
with $\kappa = \sqrt{32 \pi G}$ and one considers an effective theory defined by the Einstein-Hilbert Lagrangian
\bq
 {\mathcal L}_{\mathrm{EH}}
 & = &
 - \frac{2}{\kappa^2} 
 \sqrt{-g} R.
\eq
The field $h_{\mu\nu}$ describes a graviton.
The inverse metric $g^{\mu\nu}$ and $\sqrt{-g}$ are infinite series in $h_{\mu\nu}$, therefore
\bq
 {\mathcal L}_{\mathrm{EH}} + {\mathcal L}_{\mathrm{GF}}
 & = &
 \sum\limits_{n=2}^\infty
 {\mathcal L}^{(n)},
\eq
where ${\mathcal L}_{\mathrm{GF}}$ denotes a gauge-fixing term and 
${\cal L}^{(n)}$ contains exactly $n$ fields $h_{\mu\nu}$.
Thus the Feynman rules will give an infinite tower of vertices.
We may compute amplitudes for graviton scattering from Feynman diagrams. 
However, this is not the most efficient way to compute graviton scattering amplitudes.
We will now review alternative methods to compute the $n$-graviton tree amplitude.
One possibility comes from CHY representation.
Instead of taking one copy of $C(w, z^{(j)})$ and one copy of $E(z^{(j)},p,\eps)$ in the CHY representation,
we may take either two copies of $C(w, z^{(j)})$ or two copies of $E(z^{(j)},p,\eps)$
and ask what these formulae compute:
\bq
 m_n\left(w,\tilde{w},p\right)
 & = &
 i
 \sum\limits_{\mathrm{solutions} \; j} J\left(z^{(j)},p\right) \; C\left(w, z^{(j)}\right) \; C\left(\tilde{w}, z^{(j)}\right),
 \nonumber \\
 M_n\left(p,\eps,\tilde{\eps}\right)
 & = &
 i
 \sum\limits_{\mathrm{solutions} \; j} J\left(z^{(j)},p\right) \; E\left(z^{(j)},p,\eps\right) \; E\left(z^{(j)},p,\tilde{\eps}\right)
\eq
It turns out that $m_n(w,\tilde{w},p)$ computes 
a double-ordered amplitude of a bi-adjoint scalar theory with three-valent vertices
and that $M_n(p,\eps,\tilde{\eps})$ computes the $n$-graviton tree amplitude \cite{Cachazo:2013iea}.
The amplitude $m_n(w,\tilde{w},p)$ depends on two external orders $w$ and $\tilde{w}$,
the amplitude $M_n(p,\eps,\tilde{\eps})$ on two sets of polarisation vectors.
Gravitons are described by a product of equal-helicity polarisation vectors $\eps^\pm_\mu\eps^\pm_\nu$, 
a product of opposite helicities $\eps^\pm_\mu \eps^\mp_\nu$ corresponds to a linear combination of a dilaton 
and an anti-symmetric tensor. If all external particles are gravitons, these modes do not propagate internally.

Alternatively, we may compute the $n$-graviton amplitude from colour-kinematics duality \cite{Bern:2010ue},
by squaring the kinematical numerators:
\bq
 M_n\left(p, \eps, \tilde{\eps} \right)
 & = &
 \left(-1\right)^{n-3}
 i
 \sum\limits_{\mathrm{trivalent} \; \mathrm{graphs} \; G}
 \frac{N\left(G\right)\tilde{N}\left(G\right)}{D\left(G\right)}
\eq
A third possibility is given by the KLT relations \cite{Kawai:1985xq,Bern:1999bx,BjerrumBohr:2004wh,BjerrumBohr:2010ta,Feng:2010br,Damgaard:2012fb}:
\bq
 M_n\left(p,\eps,\tilde{\eps}\right) & = &
 - i \sum\limits_{w,\tilde{w}\in B}
 A_n^{\mathrm{YM}}\left(p,w,\eps\right) \; S_{w \tilde{w}} \; A_n^{\mathrm{YM}}\left(p,\tilde{w},\tilde{\eps}\right)
\eq
The sum is over all elements of a basis for primitive tree amplitudes in pure Yang-Mills theory.
$S_{w \tilde{w}}$ is therefore a $(n-3)! \times (n-3)!$-dimensional matrix.
This matrix is obtained as follows:
We recall that $m_n(w,\tilde{w},p)$ is the double-ordered amplitude for a bi-adjoint scalar theory with three-valent vertices.
This defines a $(n-3)! \times (n-3)!$-dimensional matrix $m_{w\tilde{w}}$ by
\bq
 m_{w\tilde{w}} & = & m_n\left(w,\tilde{w},p\right).
\eq
The matrix $m_{w\tilde{w}}$ is invertible and $S_{w\tilde{w}}$ is given by
\bq
 S_{w\tilde{w}}
 & = & \left( m^{-1} \right)_{w\tilde{w}}.
\eq
The KLT relations relate primitive tree amplitudes in pure Yang-Mills theory to graviton amplitudes.

We may now ask, if the right-hand side of the KLT relations (the one involving $A_n^{\mathrm{YM}}$) has
a generalisation from pure Yang-Mills theory to QCD.
To this aim let us consider double-ordered amplitudes $m_n^{\mathrm{flav}}\left(w,\tilde{w},p\right)$
with un-flavoured massless scalars (as before)
and flavoured scalars (massless or massive) \cite{delaCruz:2016wbr}.
We have two types of scalar vertices:
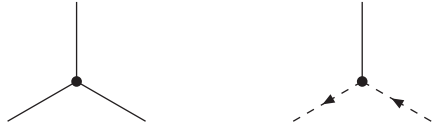
\bq
\begin{picture}(100,35)(0,50)
\Vertex(50,50){2}
\Line(50,50)(50,80)
\Line(50,50)(76,35)
\Line(50,50)(24,35)
\end{picture}
 & &
\begin{picture}(100,35)(0,50)
\Vertex(50,50){2}
\Line(50,50)(50,80)
\DashArrowLine(76,35)(50,50){3}
\DashArrowLine(50,50)(24,35){3}
\end{picture}
 \nonumber \\
\eq
Flavour is conserved.
Let us define a $N_{\mathrm{basis}} \times N_{\mathrm{basis}}$-dimensional matrix $m_{w\tilde{w}}^{\mathrm{flav}}$ by
\bq
 m_{w\tilde{w}}^{\mathrm{flav}} & = & m_n^{\mathrm{flav}}\left(w,\tilde{w},p\right).
\eq
Define further a generalised KLT matrix as the inverse of the matrix $m^{\mathrm{flav}}$:
\bq
 S^{\mathrm{flav}} & = & \left( m^{\mathrm{flav}} \right)^{-1}
\eq
We may then consider the quantity
\bq
 M_n^{\mathrm{method}\;1}\left(p,\eps,\tilde{\eps}\right) & = &
 - i \sum\limits_{w,\tilde{w}\in B}
 A_n^{\mathrm{QCD}}\left(p,w,\eps\right) \; S_{w \tilde{w}}^{\mathrm{flav}} \; A_n^{\mathrm{QCD}}\left(p,\tilde{w},\tilde{\eps}\right).
\eq
We call this relation a generalised KLT relation.

Alternatively, we might try to generalise colour-kinematics duality \cite{Chiodaroli:2013upa,Johansson:2014zca,Chiodaroli:2015wal}.
We first bring QCD amplitudes into the form
\bq
 A_n^{\mathrm{QCD}}\left(p,w,\eps\right)
 & = &
 i 
 \sum\limits_{G \in {\mathcal T}(w)}
 \frac{N\left(G\right)}{D\left(G\right)},
 \;\;\;\;\;\;\;\;\;\;\;\;
 D\left(G\right) = \prod\limits_{e \in E(G)} \left(s_e-m_e^2\right),
\eq
where the kinematical numerators $N(G)$ satisfy Jacobi-like relations, whenever the corresponding
colour factors do.
${\mathcal T}(w)$ denotes the set of all ordered tree diagrams with trivalent flavour-conserving vertices and external order $w$.
Then we set \cite{Johansson:2014zca}
\bq
 M_n^{\mathrm{method}\;2}\left(p, \eps, \tilde{\eps} \right)
 & = &
 \left(-1\right)^{n-3}
 i
 \sum\limits_{\mathrm{trivalent} \; \mathrm{graphs} \; G}
 \frac{N\left(G\right)\tilde{N}\left(G\right)}{D\left(G\right)}.
\eq
We have evidence that
\bq
 M_n^{\mathrm{method}\;1}\left(p, \eps, \tilde{\eps} \right) & = & M_n^{\mathrm{method}\;2}\left(p, \eps, \tilde{\eps} \right),
\eq
and that the quantities $M_n^{\mathrm{method}\;1/2}(p, \eps, \tilde{\eps} )$ have the properties of scattering amplitudes.
In particular they are 
invariant under (generalised) gauge transformations and the 
only poles are the single poles in the allowed factorisation channels. 
Our evidence is based on a verification of all cases with $n \le 8$ \cite{delaCruz:2016wbr}.


\section{Speculations}

We may ask what does $M_n^{\mathrm{flav}}=M_n^{\mathrm{method}\;1}=M_n^{\mathrm{method}\;2}$ compute?
To answer this question let us re-insert the coupling:
\bq
 {\mathcal M}^{\mathrm{flav}}_n\left(p,\eps,\tilde{\eps}\right)
 & = &
 \left( \frac{\kappa}{4} \right)^{n-2}
 M_n^{\mathrm{flav}}\left(p,\eps,\tilde{\eps}\right),
 \;\;\;\;\;\;\;\;\;\;\;\;
 \kappa=\sqrt{32 \pi G_N}.
\eq
In these scattering amplitudes all particles interact with gravitational strength.
Flavoured particles may be massive and hence non-relativistic.
All tree amplitudes may be computed (either by method $1$ or method $2$, where method $1$ might be more convenient
as it recycles QCD amplitudes).
In other words this defines a 
model for massive non-relativistic particles interacting only with 
gravitational strength.
This might be relevant in the discussion of dark matter.
Up to now, all evidence for dark matter is gravitational, although
most experimental searches assume additional weak-scale interactions.

However, a few comments are in order.
It is easily verified that the classical limit of massive amplitudes corresponds 
to an attractive $1/r$-potential.
Gluons and quarks have both two spin states, which we may label by $+$ and $-$.
Double copies of gluons and quarks have then four spin states:
\bq
 ++,
 \;\;\;
 +-,
 \;\;\;
 -+,
 \;\;\;
 --
\eq
In the case of a double copy of a gluon we already mentioned that
the graviton corresponds $++$ and $--$, while the states
$+-$ and $-+$ correspond to a linear combination of a dilaton and an antisymmetric tensor.
In the case of pure graviton amplitudes there is no propagation of internal $+-$ or $-+$ states.
However, this is no longer true for amplitudes with massive flavours.
As a consequence we find that in the classical limit of massive amplitudes
the effective coupling is larger by a factor $2$ due to exchange of $+-$ and $-+$ states.
It is possible to remove the dilaton and antisymmetric tensor field with the help of ghosts \cite{Johansson:2014zca}.

An important open question within this model is an explanation of the relic abundance.
This will require a non-thermal production mechanism.

Many of the developments reported in this talk on relations and representations of amplitudes 
have been inspired by string theory or supersymmetry.
However it is worth pointing out that the results do not rely on string theory nor supersymmetry.


\section{Conclusions}

In this talk we considered relations and representations of primitive QCD tree amplitudes.
We focused on how these relations and representations generalise from pure Yang-Mills theory to QCD.
The generalisation of the KLT relations from pure Yang-Mills to QCD includes the case of massive quarks.
On the gravity side we then obtain hypothetical particles interacting with gravitational strength,
which can be massive and non-relativistic.
This might be relevant in the discussion of dark matter and is certainly worth to study in more detail.


\bibliography{/home/stefanw/notes/biblio}
\bibliographystyle{/home/stefanw/latex-style/h-physrev5}

\end{document}